\documentclass[12pt]{article}
\usepackage{times}
\usepackage{geometry}
\usepackage[utf8]{inputenc}
\usepackage{enumitem,amssymb}
\usepackage{ragged2e}
\usepackage[]{natbib}
\bibpunct{(}{)}{;}{a}{}{,}
\usepackage{wrapfig}

\usepackage{myusepack}
\newlist{thematic}{itemize}{8}
\setlist[thematic]{label=$\square$}
\usepackage{pifont}
\usepackage{graphicx}
\newcommand{\lya}{Ly$\alpha$}

\newcommand{\hb}{H$\beta$}
\newcommand{\fe}{$f^{e}_{LyC}$}
\newcommand{\fen}{$f^{e}_{900}$}

\newcommand{\HST}{{\em HST}\/}
\newcommand{\hst}{{\em HST}\/}

\newcommand{\JWST}{{\em JWST}\/}
\newcommand{\jwst}{{\em JWST}\/}

\newcommand{\flux}{ergs cm$^{-2}$ s$^{-1}$ \AA$^{-1}$ }
\newcommand{\cmtwosa}{cm$^2\;$s$\;$\AA}
\begin{document}
\pagestyle{empty}
\huge
Astro2020 Science White Paper  \hspace{\fill} \linebreak

\noindent Lyman continuum observations across cosmic time: recent developments, future requirements \hspace{\fill} 
\normalsize

\noindent \textbf{Thematic Areas:} \hspace*{31pt} $\square$    Galaxy Evolution  \hspace*{45pt} $\square$ Cosmology and Fundamental Physics  \linebreak
  
\noindent \textbf{Principal Author:}

\noindent Stephan R. McCandliss	\hspace{\fill} 
 \linebreak						
 Johns Hopkins University,
 Department of Physics and Astronomy, \\
 Center for Astrophysical Sciences, 3400 North Charles Street, Baltimore, MD 21218 \hspace{\fill} 
 \linebreak
stephan.mccandliss@jhu.edu,
410-516-5272 \hspace{\fill}
 \linebreak
 
\noindent \textbf{Co-authors:} 
  \linebreak
 \noindent Daniela Calzetti (astro.umass.edu),
  Henry C. Ferguson (stsci.edu),
  Steven Finkelstein (astro.as.utexas.edu),
  Brian T. Fleming (colorado.edu),
  Kevin France (colorado.edu),
  Matthew Hayes (astro.su.se),  
  Timothy Heckman (jhu.edu),
  Alaina Henry (stsci.edu),
  Akio K. Inoue (aoni.waseda.jp),
  Anne Jaskot (astro.umass.edu),
  Claus Leitherer, (stsci.edu),
  Sally Oey (umich.edu),
  John O'Meara (keck.hawaii.edu),
  Marc Postman (stsci.edu),
  Laura Prichard (stsci.edu)
  Swara Ravindranath (stsci.edu),
  Jane Rigby (nasa.gov),
  Claudia Scarlata (astro.umn.edu),
  Daniel Schaerer (unige.ch),
  Alice  Shapley (astro.ucla.edu),
  Eros Vanzella (inaf.it) \\
  
\noindent \textbf{Abstract:}  Quantifying the physical conditions that allow radiation emitted shortward of the hydrogen ionization edge at 911.7 \AA\ to escape the first collapsed objects and ultimately reionize the universe is a  compelling problem for astrophysics. The escape of LyC emission from star-forming galaxies and active galactic nuclei is intimately tied to the emergence and sustenance of the metagalactic ionizing background (MIB) that pervades the universe to the present day and in turn is tied to the emergence of structure at all epochs.  The James Webb Space Telescope (\jwst) was built in part to search for the source(s) responsible for reionization, but it cannot observe LyC escape directly, because of the progressive increase in the mean transmission of the intergalactic medium towards the epoch of reionization.  Remarkable progress has been made to date in directly detecting LyC leaking from star-forming galaxies using space-based and the ground-based observatories, but there remain significant gaps in our redshift coverage of the phenomenon.  Ongoing projects to measure LyC escape at low- and intermediate-$z$ will provide guidance to \jwst\ investigations by analyzing the robustness of a set of proposed LyC escape proxies, and also provide a closeup examination of the physical conditions that favor LyC escape.  However, currently available facilities are inadequate for deeply probing  LyC escape at the faint end of the galaxy luminosity function.  Doing so will require facilities that can detect LyC emission in the restframe to limiting magnitudes approaching 28 $ < m^*_{(1+z)900} <$ 32 for $M^*_{(1+z)1500}$ galaxies.    The goal of acquiring statistically robust samples for determining LyC luminosity functions across cosmic time will require multi-object spectroscopy from spacebased flagship class and groundbased ELT class telescopes along with ancillary panchromatic imaging and spectroscopy spanning the far-UV to the mid-IR.



\pagebreak
\pagestyle{plain}
\setcounter{page}{1}
\section{Introduction} \vspace{-.1in}
	
Quantifying the physical conditions that allow radiation emitted shortward of the hydrogen ionization edge at 911.7 \AA\ (the Lyman Continuum; LyC) to escape from star-forming galaxies and active galactic nuclei (AGN) throughout cosmic time is one of the most compelling questions for astrophysics.  LyC escape governs the timescale for the phase transition in the intergalactic medium (IGM) from a radiation-bounded to a density-bounded state that occurred when the universe was $\lesssim$ 1 Gyr old; a period known as the Epoch of Reionization (EoR).  It resulted in the emergence and sustenance of the metagalactic ionizing background (MIB) that pervades the universe to the present.  Our goal is to determine the luminosity functions of LyC emitters across cosmic time. %
	
Reionization requires there to be at least one ionizing photon for every baryon \citep{Madau:1999}.  The ionizing photon production rate is expressed as,
\begin{equation}
\dot{N}_{gal} = f^e_{LyC} \xi_{ion} \rho_{SFR},
\end{equation}

\noindent where $f^e_{LyC}$  is the average fraction of LyC photons that escape from galaxies, $\xi_{ion}$  is the average production rate of LyC photons per unit star-formation rate density (SFRD)  (the ionizing emissivity).  The SFRD density, $\rho_{SFR}$, is well characterized quantity \citep[][references therein]{Madau:2014}, however, the product of $f^e_{LyC}\xi_{ion}$ is less well constrained.  How \fe\ and $\xi_{ion}$ evolve with redshift are major systemic unknowns \citep{Finkelstein:2019, Ellis:2014, Haardt:2012}, impeding our understanding of the thermal history of the IGM \citep{Puchwein:2019}, and quantifying the impact of the MIB on the formation of structures at all epochs.   %
	
At low redshifts, the intensity of the MIB has been tied to a lower than expected  \ion{H}{1} column density distribution of \lya\ forest absorbers, (12.5 $< \log{[N_{HI}(cm2)]} <$ 14.5).  This has been termed an ionizing photon underproduction crisis, where the paucity in the number of forest absorbers \citep{Danforth:2016} requires an MIB (and consequently \fe) to be $\approx$ 2 to 5 times higher than theoretical estimates \citep[c.f.][]{Haardt:2012, Kollmeier:2014, Shull:2015, Puchwein:2015, Khaire:2015, Gaikwad:2017, Tonnesen:2017}.  At higher redshifts, it is an open question whether the primary source of the MIB was stars or AGN \citep{Madau:2015}, although recent observations of Lyman Break Galaxies (LBG)  at $z \approx$ 3 \citep{Steidel:2018}  indicate that the ionizing emissivity from star-forming galaxies exceeded that of quasi-stellar objects by $\geq 50$\%.  %

\begin{figure}
\includegraphics[width=0.48\textwidth]{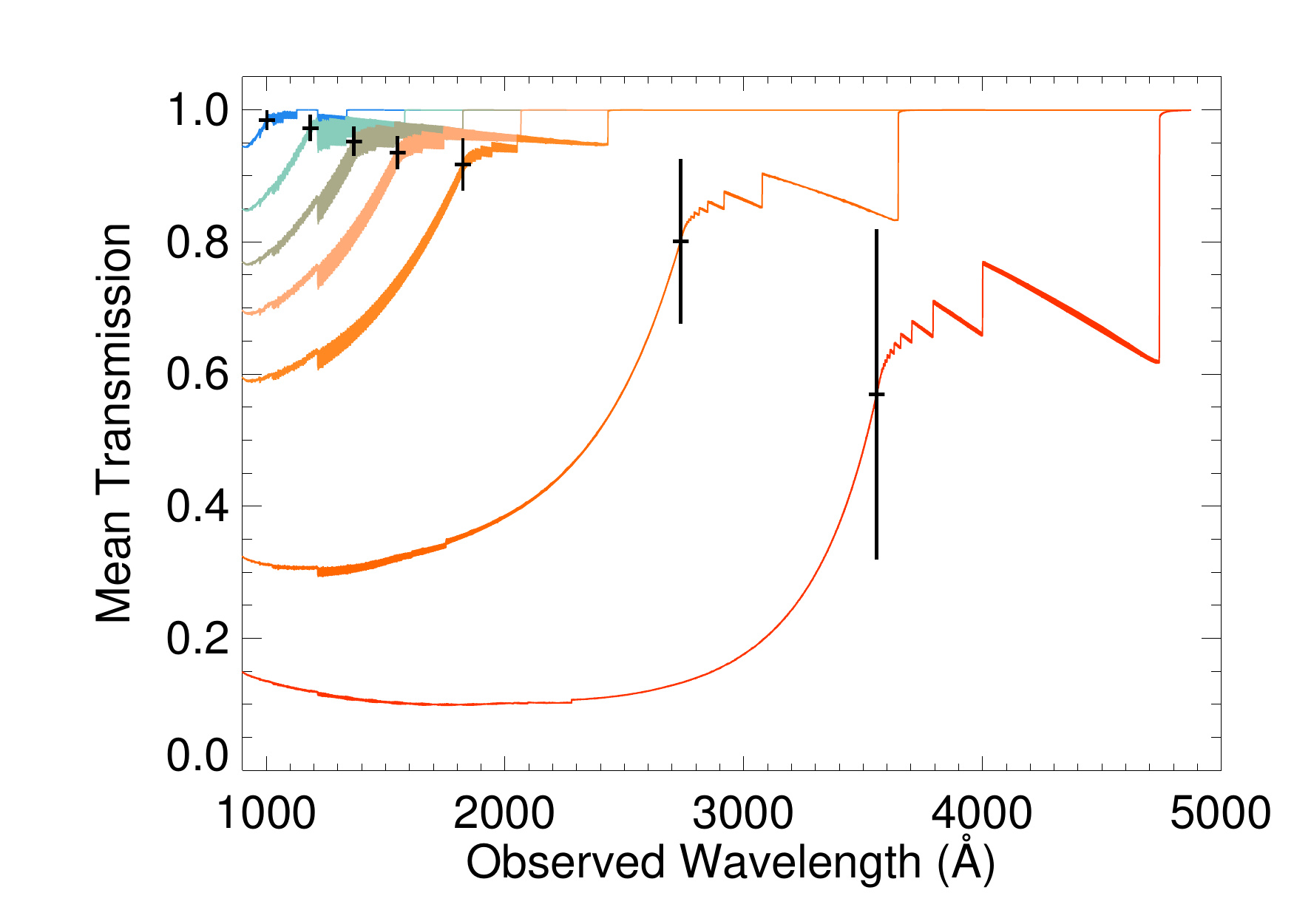} \hspace{\fill}
\includegraphics[width=0.42\textwidth]{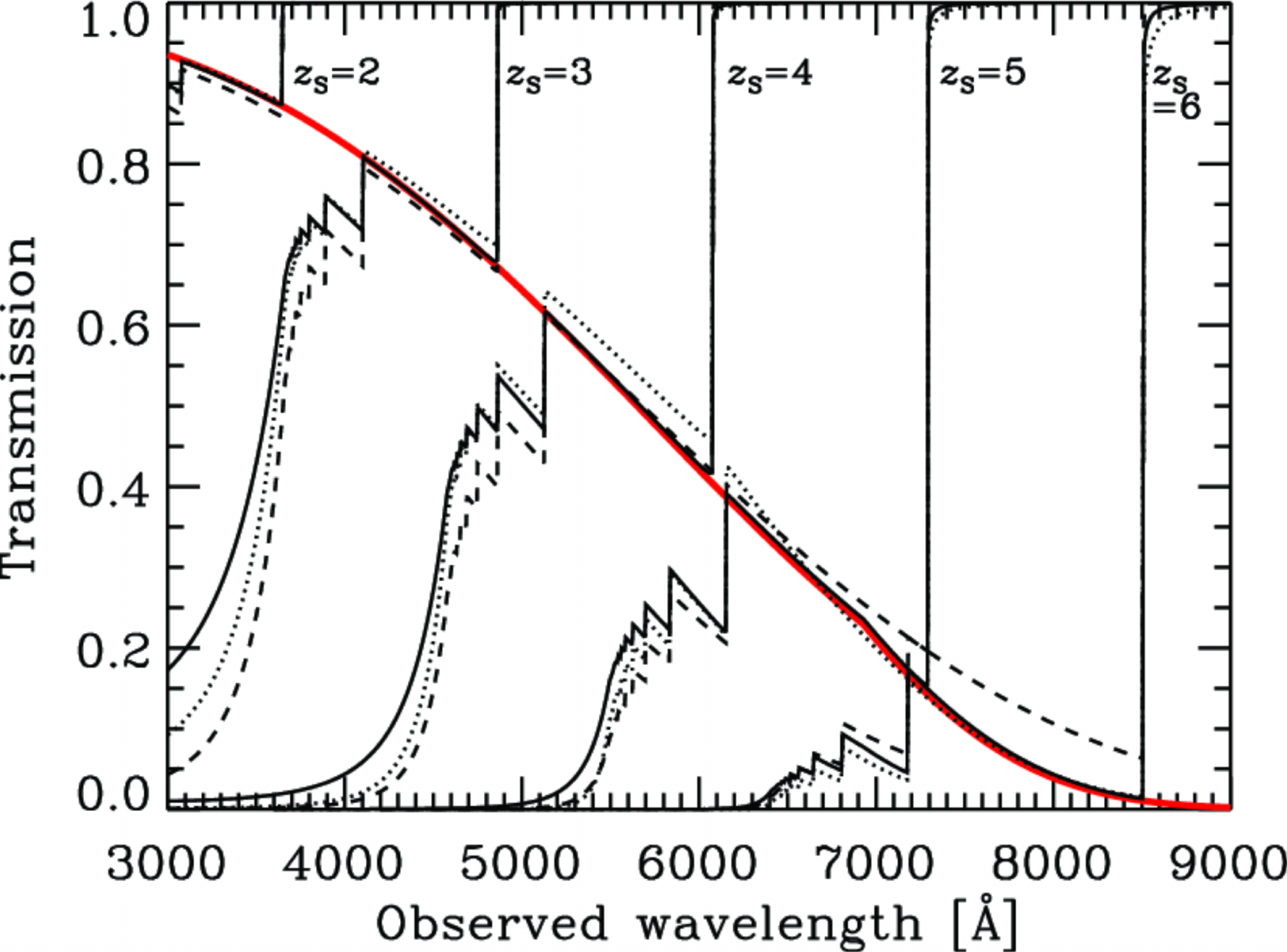}
\caption{\small Left - Mean IGM transmission  for  $z$ = (0.1, 0.3, 0.5, 0.7, 1.0, 2.0, 2.9) \citep{McCandliss:2017}. Vertical bars mark the Lyman edge and indicate the level of expected variation found in a Monte Carlo study of IGM transmission by \citet{Inoue:2008}.  Right - same for $z$ = (2, 3, 4, 5, 6) \citep{Inoue:2014}.  The red line on the right marks the mean transmission at \lya.  } \label{f1}
\end{figure}

During the EoR small galaxies may plausibly dominate the ionizing radiation budget provided: the faint end slope of the galaxy luminosity function was steep, that it extended to absolute UV magnitudes $M_{1500}$ $\sim$ -13 mag, and that 5\% $<$ \fe\ $<$ 40\% \citep[c.f.][and references therein]{Bouwens:2015,  Khaire:2016, Finkelstein:2019}.  The James Webb Space Telescope (\JWST), built in part to search for the source(s) of the EoR, can only do so indirectly.  The monotonic increase with redshift in the density of Lyman limit systems (LLS) -- those discrete clouds in the IGM having  $\log(N_{HI}(cm^{-2})) > $ 17.2 --  produces an severe drop in the mean transmission of the IGM towards $z \sim$ 6 \citep{Madau:1995, Inoue:2014, McCandliss:2017} as shown in Figure ~\ref{f1}.  Projects seeking to directly quantify  $f^e_{LyC}$ and  $\xi_{ion}$ at low- and intermediate-$z$ require space- and ground-based facilities with extremely sensitive UV capabilities.  %

Here we review confirmed detections to date, present the potential of LyC escape proxies and needs for ancillary observations in the visual and IR, discuss the currently approved direct detection projects, summarize key theoretical questions, and describe future requirements for fully characterizing the ionizing radiation luminosity function across cosmic time. %

\section{Recent Progress in LyC Direct Detections}

\begin{wrapfigure}{R}{.4\textwidth}	\vspace{-1.6in} 	
\includegraphics[width=0.5\textwidth]{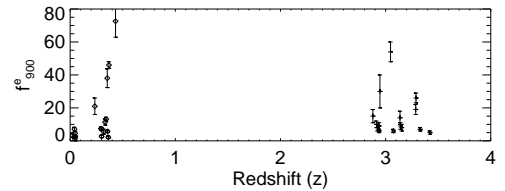}  \vspace{-.35in}
\caption{\small Confirmed \fen\ measurements and redshifts from the literature. } \label{f2}
\end{wrapfigure}

	Remarkable progress has been made in obtaining statistically significant spectroscopic detections of LyC leaking from star-forming galaxies both in the far-UV around $z \sim$ 0.3, using Cosmic Origins Spectrograph (COS)  on \hst, and from the ground in $U$ and $B$ bands around $z \sim$ 3.  They are shown in Figure~\ref{f2}.  The critical redshift range between 0.3$\lesssim z \lesssim$ 3 is an observational desert ripe for exploration by new high sensitivity instruments.
	

	{\bf Spectroscopic detections to date at low redshift:}  Prior to 2016, only three $z \sim$  0 galaxies had LyC detections, all with \fen\ $ \lesssim $3\% \citep{Leitet:2013,  Borthakur:2014}.  Earlier low-$z$ work had produced only upper limits of \fen\ $<$ 10 \%  for nine galaxies \citep{Leitherer:1995, Leitet:2013}.  More recently \citet{Leitherer:2016} reported on three detections of objects with 0.041 $< z <$  0.048 with 2.5 $\le$ \fen\  $<$ 7, and \citet{Izotov:2018a, Izotov:2018b} have found 12 additional LyC emitters with  2.5 $\le$ \fen\  $\le$ 72 in between 0.3 $< z <$  0.4.  The latter sample are the compact green-peas (GPs) galaxies that reside on the upper end of the ``BPT'' star-forming galaxy sequence \citep{Baldwin:1981}.  They were selected for having high \hb\ equivalent widths (EW(\hb) $>$ 200 \AA) and  [\ion{O}{3}]/[\ion{O}{2}] $>$ 5 in Sloan Digital Sky Survey (SDSS) spectra.  They are also characterized by relatively narrow double peaked \lya\ profiles in COS medium resolution spectra. Spectroscopy at low-$z$ suffers little IGM attenuation and allows closeup viewing of leaky regions. 
	
	{\bf Detections to date at intermediate redshift: }  Early follow-up \HST\ observations of ground-based $z \sim$ 3 LyC detections  suffered from contamination by foreground galaxies \citep[e.g.][]{Mostardi:2015, Siana:2015}, but recently, LyC emission has been detected in $\sim$ 30 individual galaxies  \citep{Mostardi:2015, Shapley:2016, Vanzella:2016, Vanzella:2018, Fletcher:2018, Steidel:2018}, as well as in stacked samples \citep{Marchi:2018, Steidel:2018}.  The derived \fen\ require large corrections for the transmission of the intervening IGM, but identifying ``lucky-sightlines'' with 2$\sigma$ high IGM transmissions will mitigate this issue.

	\section{Indirect LyC Proxies, Ancillary Data, and Analogs}
	
The opacity of the IGM at $z >$ 6 will cause \jwst\ to rely upon indirect diagnostics of LyC escape to identify the source(s) of reionization.   We review candidate LyC proxies that can be quantified at low-$z$.  
There is substantial scatter in these indicators.   It is critical to expand on these proxy investigations to quantify the emergence and sustenance of the MIB from the EoR to the present. 	
		
	
		{\bf  \lya\ emission} is the most commonly suggested proxy for LyC escape.  The high EW of the line makes it more readily detectable against weaker continuum  at high-$z$.  Escaping \lya\ is a somewhat paradoxical  LyC escape indicator as its emission is a consequence of H ionized by LyC photons that do not escape. Nevertheless, observations indicate strong \lya\ emission correlates with LyC escape in both $z \sim$  3 stacked samples \citep[e.g.][]{Steidel:2018, Marchi:2018} and among individual low-z LyC emitters \citep{Verhamme:2017};  however, the sample of strong LyC emitters is still small.  \citet{Verhamme:2015} and \citet{Dijkstra:2016} have calculated that LyC emitting galaxies will have high \lya\ escape fractions, with telltale closely-spaced, double-peaked spectral profiles spanning the systemic velocity; a consequence of reduced scattering in low column density \ion{H}{1} gas. The usefulness of \lya\ emission at $z >$ 6 is in some doubt as the IGM will absorb the blue portion of double-peaked \lya\ profiles, although the velocity offset of the red peak may be less affected and could constrain \fe\ \citep[e.g.][]{Yang:2017}.
		  
		  {\bf  [\ion{O}{3}] $\lambda$5007]/[\ion{O}{2}] $\lambda$3727} has become the most successful predictor of LyC emission at low- and intermediate-$z$ \citep{Izotov:2016, Izotov:2018b, Fletcher:2018, Vanzella:2016}.  This diagnostic measures the ionization parameter for photoionized gas.  Extreme values are not necessarily diagnostic of LyC escape \citep[e.g.][]{Stasinska:2015, Jaskot:2013}.   However, extreme [\ion{O}{3}]/[\ion{O}{2}]  appears to be linked to optically thin channels as evidenced by a strong correlation with \lya\ peak separation, and hence, specific gas geometries favorable to LyC escape (Jaskot et al. 2019, submitted ApJ). 


		
		{\bf \ion{Mg}{2} \ion{Mg}{2} $\lambda\lambda$ 2796, 2803 emission} has been reported to have a strong correlation with \lya\ emission  by \citet{Henry:2018}  in observations at low-$z$ of extreme emission line galaxies, suggesting a potential a new proxy for LyC radiation that can be accessed by \jwst\ at $z \gtrsim$ 6 where they will appear at wavelengths $\lambda \gtrsim$ 1.96 $\mu m$. 
				
		{\bf High  star-formation surface density ($\Sigma_{SFR}$)} may produce concentrated feedback on galaxies perhaps efficiently clearing holes for LyC photons to escape \citep{Heckman:2001, Clarke:2002}.    LyC observations \citep{Verhamme:2017} and stacked spectra \citep{Marchi:2018} suggest that \fe\ may indeed increase with $\Sigma_{SFR}$, although the data are sparse.  \citet{Sharma:2017} used a simple model with fesc = 0.2 and $\Sigma_{SFR} >$ 0.1 M$\odot$ yr$^{-1}$ kpc$^{-2}$ to predict that galaxies have progressively higher \fe\ toward high redshift, however, Figure~\ref{f2} does not support this contention.  
		
	{\bf Absorption line-core residual intensity; \lya\ blue wing; Weak [\ion{S}{2}]: } \citet[][and references therein]{Alexandroff:2015} identified 3 indirect proxies for LyC escape in Lyman Break  Analogs (LBAs): (1) the residual intensity in the cores of saturated interstellar low-ionization absorption lines, indicating incomplete covering by that gas in the galaxy, (2) the relative strength of blueshifted \lya\ line emission, indicating the existence of holes in the \ion{H}{1} on the front-side of the galaxy outflow, and (3)  weak visual [\ion{S}{2}] emission, indicative of a density-bounded \ion{H}{2} region. 

		{\bf Ancillary Data Sets:}  Observations of LyC are necessary but not sufficient for understanding the physical conditions that favor LyC escape.   Spectroscopy in the UV-visual restframe provides diagnostics for  attenuation by dust, metallicity, radiation harness, and the total production rate of ionizing photons \citep{Osterbrock:2006}.  Far-UV spectra at medium resolution ($R \sim $ 20,000) around restframe \lya\ probe emission and differences in absorption line velocity-fields and covering factions for \ion{H}{1}, and low- and high-ionization metals species \citep{Henry:2015, Reddy:2016}.  Lower resolution spectra probe the LyC along with far-UV dust attenuation from determinations of the  logarithmic slope, $\beta$\footnote{$f_{\lambda} = f_o (\lambda/\lambda_{o})^{\beta}$.  Attenuated star-forming galaxies have $\beta \gtrsim -2 $ }.  High resolution narrow- and broad-band photometry from the far-UV to the mid-IR are used to assess galaxy morphology (geometry), determine extension of \lya\ \citep[][and citations to]{Hayes:2014},  provide color information for candidate target selection  \citep[e.g.][]{Nestor:2013, Inoue:2011}, and assessment of $\Sigma_{SFR}$.

{\bf On the existence of low- and intermediate-$z$  analogs of  high-$z$  reionizers:}  There is a general overlap between GP, LBA  at low-$z$ and LBG at intermediate-$z$.   They all populate the upper end of the BPT diagram.  LBG have been found at 11 $\gtrsim z$ \citep{Vulcani:2017, Livermore:2017}.   GP are compact and resemble high-redshift galaxies through their low metallicities, low extinction, high UV luminosities, and enormous $\Sigma_{SFR}$ \citep{Izotov:2011}, while LBAs and LBGs have higher masses and metallicities \citep[c.f.][]{Izotov:2018b, Strom:2018, Heckman:2011, Overzier:2009}.  Some LBA  contain a dominant central object with $\sim$ 10$^9$ M$_{\odot}$ within a 100 pc radius, but without significant AGN activity from an accreting supermassive blackhole (BH), perhaps because they are dominated by supernova outflow.   There is speculation that GP $\rightarrow$ LBG may represent a continuum of pre-QSO type objects at high-$z$.  Measuring  LyC escape in GP in comparison to LBA will inform us of the expected trends with mass and metallicity in the EoR.


	\section{Ongoing direct detection projects at  low-$z$}
		{\bf Low-Redshift Lyman Continuum Survey:}   \citet[][\hst\ - 15626,]{Jaskot:2018} has been granted a 134 orbit program using the CENWAV800 mode on COS \citep{Redwine:2016} to obtain the first large statistical sample of LyC measurements for  66 $z \sim$ 0.3 star-forming galaxies.  It is expected to establish limits on \fen\ to  5\%. The program will examine five proxies of LyC escape, which are accessible to JWST at $z >$ 6: 1) high [\ion{O}{3}]/[\ion{O}{2}], 2) strong, narrow \lya\ emission, 3) weak low-ionization UV absorption lines, 4) high star-formation surface densities, and 5) reduced \hb\ EW.  The sample will discriminate between the different proposed diagnostics, quantify LyC scaling relations and their scatter, and provide fundamental statistics on the LyC emission within different selected populations.  Theoretical models will provide understanding of indirect LyC diagnostics by simulating the expected spectra from LyC emitters with realistic galaxy geometries. This program will reveal how LyC emission correlates with galaxy physical properties, thereby providing crucial information for JWST studies of reionization.
			
		{\bf Supernova-remnants/Proxies for Reionization/ and Integrated Testbed Experiment:}  
		\citep[SPRITE,][]{Fleming:2019} is a 6U CubeSat that seeks to determine how galaxies provide ionizing radiation to the IGM.  It is expected to have an effective area and spectral resolution, $\sim$ 10 - 15 cm$^{2}$ and $R \approx$ 1500 respectively, and to reach a background limited flux of $\approx$ 10$^{-17}$ \flux similar to that achievable with COS.  The anticipated 7 to 12 month mission lifetime will allow SPRITE to devote over $\sim$ 150 days to directly measure the ionizing spectrum of 100 $z \lesssim$ 0.3 galaxies and AGN  with exposure durations of $\sim$ 60ks thereby  establishing limits on \fen\  $\lesssim$ 2\%.  This sample  will surpass in number and precision all ionizing escape measurements to date and provide critical interpretive tools for core \JWST\ science.  SPRITE is planning for a late-2021/early-2022 launch. 

\section{Key Theoretical Questions  -- Observational Requirements }

\begin{wrapfigure}{R}{.5\textwidth} \vspace{-.35in}	
\includegraphics[angle=90,width=.5\textwidth]{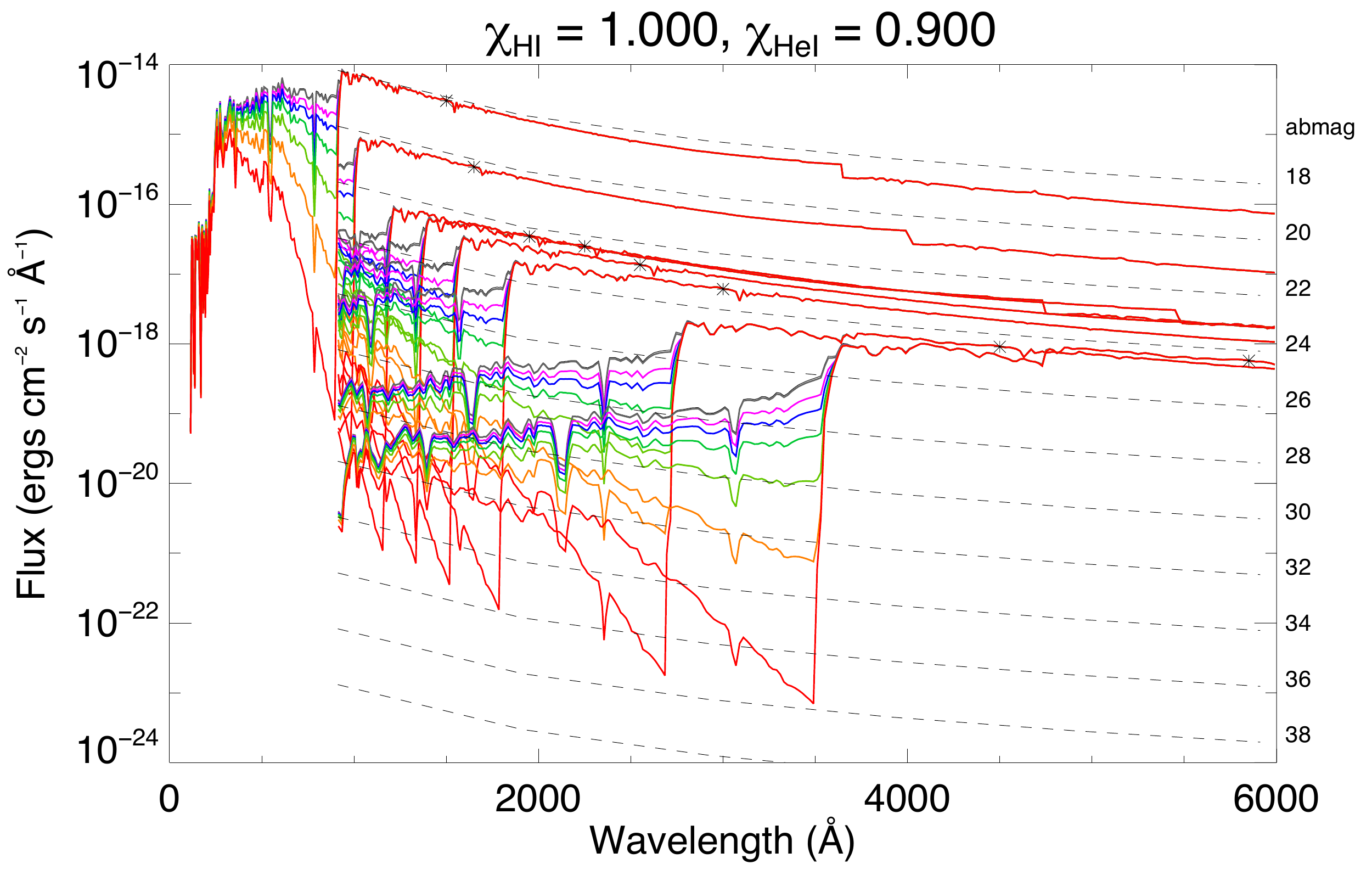}
\vspace{-.4in}
\caption{\small Log scaled model galaxy fluxes for $z$ = (0, 0.1, 0.3, 0.5, 0.7, 1.0, 2.0, 2.9) with $f^e_{900}$ (0.000, 0.003, 0.041, 0.166, 0.364, 0.566, 0.945)  in (red, orange, light green, green, blue, violet, and grey) respectively.   Contours of constant abmag appear as dashed lines.  Asterisks mark fluxes for $m^*_{(1+z)1500}$.   \label{f3}}
\vspace{-.2in}
\end{wrapfigure} %

Understanding  the diversity of  \fe\  at low- and intermediate-$z$ and establishing connections to the EoR will require sophisticated radiative hydrodynamical simulations that include the physics of dark matter halos, gas dynamics and kinematics, star formation history evolution from high- to low-$z$, BH growth, radiative and mechanical feedback, and  most importantly line-of-sight geometry variance \citep[c.f.][]{Zackrisson:2017, Cen:2015, Yajima:2014}.    Key questions that require theoretical and observational guidance include:

		
		\noindent {\bf  How do $\xi_{ion}$ and $f^e_{LyC}$  vary as a function of  mass, metallicity, star formation history, and redshift?}   We know stellar population models aren't producing enough hard photons because of their inability to reproduce nebular \ion{He}{2} \citep{Schaerer:2019, Berg:2018}, raising questions regarding our assumptions of massive star IMFs, and the presence of x-ray binaries.
				
		\noindent {\bf What local physical conditions regulate LyC escape?}  Here high spectral/spatial resolution on star-forming clusters at 30-40 pc scales can quantify neutral gas covering fractions,  and explore  radiative and mechanical feedback effects.
		
	{\bf Observational Requirements:}   \citet{Finkelstein:2019} have recently explored various scenarios for the EoR wherein the average galaxy \fe\ $ < $ 5\%.  Such low LyC escape scenarios require: a low end absolute magnitude limit
of -13 $ < M_{UV,lim} < $ -11; galaxies becoming more efficient producers of ionizing photons towards lower magnitudes and higher redshift; and  a non-dominant but significant contribution to LyC escape by AGN at $z \lesssim$ 7.  These findings provide impetus to extend constraints on $\xi_{ion}$ and $f^e_{LyC}$ to objects at the faint end of the luminosity function.
		
	Reaching such faint objects require much fainter background limits than available to the currently approved programs. \citet{McCandliss:2017} have determined LyC detection requirements and the areal densities of candidate emitters as a functions of redshift.  They were derived from a 10 Myr  Starburst99 spectral synthesis model \citep[][Geneva tracks with rotation at 40\%  of breakup velocity]{Leitherer:2014}, and the far-UV luminosity functions of \citet{Arnouts:2005}. The corresponding SEDs are shown in Figure~\ref{f3}.  They feature the ``drop-in'' of flux shortward of the edge with \ion{H}{1} optical depth $\propto (\lambda/912)^3$; an empirical measure of $\xi_{ion}$.  The number of $M^*_{1500}$ galaxies per square degree range from $\sim$ 10$^2$ at $z$ = 0.1, to $\sim$ 10$^4$ by $z$ = 3.  Their redshifted apparent UV magnitudes are 20 $< m^*_{(1+z)1500} <$ 25, and for \fen\ $ \approx $ 0.3 \%, the over-the-edge magnitude is 28 $ < m^*_{(1+z)900} <$ 32.   Reaching these depths (5$\sigma$ detections) requires a product of effective area $A^{eff}_{\lambda}$, observing time $\Delta T$, and bandwidth $\Delta \lambda$ of 4 $\times$ 10$^8$ $ < A^{eff}_{\lambda}~\Delta T~\Delta \lambda <$ 10$^{11}$ \cmtwosa .

{\bf We conclude that acquiring  statistically relevant samples for determining LyC luminosity functions across cosmic time will require multi-object spectroscopy from spacebased flagship class and groundbased ELT class telescopes along with ancillary  panchromatic imaging and spectroscopy spanning the far-UV to the mid-IR.}


\pagebreak

\small
\bibliography{ms.bbl}

\end{document}